\documentclass[12pt]{article}
\usepackage{amsmath}
\usepackage{amssymb}

\input{tcilatex}

\begin{document}

\bigskip\ 

\bigskip\ 

\begin{center}
\textbf{LINEARIZED FIVE DIMENSIONAL }

\smallskip\ 

\textbf{KALUZA-KLEIN THEORY AS A GAUGE THEORY}

\textbf{\ }

\smallskip\ 

G. Atondo-Rubio, J. A. Nieto \footnote{%
nieto@uas.uasnet.mx}, L. Ruiz and J. Silvas

\smallskip

\textit{Facultad de Ciencias F\'{\i}sico-Matem\'{a}ticas, Universidad Aut%
\'{o}noma}

\textit{de Sinaloa, C.P. 80000, Culiac\'{a}n, Sinaloa, M\'{e}xico}

\bigskip\ 

\bigskip\ 

\textbf{Abstract}
\end{center}

We develop a linearized five dimensional Kaluza-Klein theory as a gauge
theory. By perturbing the metric around flat and the De Sitter backgrounds,
we first discuss linearized gravity as a gauge theory in any dimension. In
the particular case of five dimensions, we show that using the Kaluza-Klein
mechanism, the field equations of our approach implies both linearized gauge
gravity and Maxwell theory in flat and the De Sitter scenarios. As a
possible further development of our formalism, we also discuss an
application in the context of a gravitational polarization scattering by
means of the analogue of the Mueller matrix in optical polarization algebra.
We argue that this application can be of particular interest in
gravitational waves experiments.

\bigskip\ 

\bigskip\ 

\bigskip\ 

\bigskip\ 


Pacs numbers: 04.50.+h, 04.30.-w, 98.80.-k, 42.15.-i

September, 2006

\newpage \noindent \textbf{1.- Introduction}

\smallskip\ 

It is known that linearized gravity can be considered as a gauge theory [1].
In this context, one may be interested in the idea of a unified theory of
linearized gravity and Maxwell theory. This idea is, however, no completely
new since in fact the quest of a unified theory of gravity and
electromagnetism has a long history [2]. One can mention, for instance, the
five dimensional Kaluza-Klein theory [3] which is perhaps one of the most
interesting proposals. The central idea in this case is to incorporate
electromagnetism in a geometrical five dimensional gravitational scenario.
The gauge properties arise as a result of broken general covariance via a
mechanism called spontaneous compactification. Symbolically, one may
describe this process through the transition $M^{5}\rightarrow M^{4}\times
S^{1},$ where $M^{5}$ and $M^{4}$ are five and four dimensional manifolds
respectively and $S^{1}$ is a circle. Thus, after compactification the
fiber-bundle $M^{4}\times S^{1}$ describes the Kaluza-Klein scenario. Let us
picture this attempt of unification as 
\begin{equation}
em\rightarrow g,  \tag{1}
\end{equation}%
where $em$ means electromagnetism and $g$ gravity.

In the case of linearized gravity theory the scenario looks different
because it can be understood as a gauge theory rather than a pure
geometrical structure. Therefore, a unified theory in this case may be
understood as an idea to incorporate linearized gravity in a gauge Maxwell
context. In other words, one may start from the beginning with a generalized
fiber-bundle $M^{4}\times B$, with $B$ as a properly chosen compact space.
Thus, we have that this case can be summarized by the heuristic picture

\begin{equation}
em\leftarrow g.  \tag{2}
\end{equation}

Our aim in this paper is to combine the two scenarios (1) and (2) in the form

\begin{equation}
em\longleftrightarrow g.  \tag{3}
\end{equation}%
Specifically, we start with linearized gravity in five dimensions and apply
the Kaluza-Klein compactification mechanism. We probe that the resultant
theory can be understood as a gauge theory of linearized gravity in five
dimensions. Furthermore, we show that by using this strategy one can derive
a unified theory of gravity and electromagnetism with a generalized gauge
field strength structure. As an advantage of our formalism, we outline the
possibility that optical techniques can be applied to both gravity and
electromagnetic radiation in a unified context. Thus, we argue that our
results may be of particular interest in the gravitational waves detection.

Technically this article is organized as follows. In section 2, we develop
linearized gravity in any dimension. In section 3, we discuss linearized
gravity in a 5-dimensional Kaluza-Klein theory and in section 4 we
generalize our procedure to a De Sitter background. In section 5, we outline
a possible application of our formalism of unified framework of
electromagnetic and gravitational radiation on an optical geometry via the
Mueller matrix. In appendix A we briefly review the De Sitter background
theory and in the appendix B we present a generalization to any dimension of
the Novello and Neves work [4].

\bigskip\ 

\noindent \textbf{2.- Linearized gravity in any dimension}

\smallskip\ 

Let us consider a $1+d$-dimensional manifold $M^{1+d},$ with associated
metric $\gamma _{AB}(x^{C}).$ We shall assume that $\gamma _{AB}$ can be
written in the form

\begin{equation}
\gamma _{AB}=\eta _{AB}+h_{AB},  \tag{4}
\end{equation}%
where $\eta _{AB}=diag(-1,1,...,1)$ and $h_{AB}(x^{C})$ is a small
perturbation, that is

\begin{equation}
\left\vert h_{AB}\right\vert <<1.  \tag{5}
\end{equation}%
To first order in $h_{AB}$, the inverse of the metric $\gamma _{AB}$ becomes

\begin{equation}
\gamma ^{AB}=\eta ^{AB}-h^{AB}.  \tag{6}
\end{equation}%
Using (4) and (6) we find that the Christoffel symbol and Riemann curvature
tensor are

\begin{equation}
\Gamma _{CD}^{A}=\frac{1}{2}\eta ^{AB}(h_{BC,D}+h_{BD,C}-h_{CD,B}),  \tag{7}
\end{equation}%
and

\begin{equation}
R_{ABCD}=\partial _{A}\mathcal{F}_{CDB}-\partial _{B}\mathcal{F}_{CDA}, 
\tag{8}
\end{equation}%
respectively. Here, the symbol $\mathcal{F}_{CDB}$ means

\begin{equation}
\mathcal{F}_{CDB}=\frac{1}{2}(h_{BC,D}-h_{BD,C}).  \tag{9}
\end{equation}%
Observe that $\mathcal{F}_{CDB}$ is antisymmetric in the indices $C$ and $D.$

In terms of the quantity $\mathcal{F}_{A}$ defined by

\begin{equation}
\mathcal{F}_{A}=\eta ^{CB}\mathcal{F}_{ACB},  \tag{10}
\end{equation}%
and the symbol $\mathcal{F}_{ADB}$, the Ricci tensor $R_{BD}$ reads

\begin{equation}
R_{BD}=\partial ^{A}\mathcal{F}_{ADB}+\partial _{B}\mathcal{F}_{D}.  \tag{11}
\end{equation}%
Thus, we get that the Ricci scalar $R$ is given by

\begin{equation}
R=2\partial ^{A}\mathcal{F}_{A}.  \tag{12}
\end{equation}

Substituting (11) and (12)\ into the Einstein weak field equations in $1+d$
dimensions

\begin{equation}
R_{BD}-\frac{1}{2}\eta _{BD}R=\frac{8\pi G_{1+d}}{c^{2}}T_{BD,}  \tag{13}
\end{equation}%
we find

\begin{equation}
\partial ^{A}\mathcal{F}_{ADB}+\partial _{B}\mathcal{F}_{D}-\eta
_{BD}\partial ^{A}\mathcal{F}_{A}=\frac{8\pi G_{1+d}}{c^{2}}T_{BD},  \tag{14}
\end{equation}%
where $G_{1+d}$ is the Newton gravitational constant in $1+d$ dimensions.

Let us now define the symbol%
\begin{equation}
F_{ADB}\equiv \mathcal{F}_{ADB}+\eta _{BA}\mathcal{F}_{D}-\eta _{BD}\mathcal{%
F}_{A},  \tag{15}
\end{equation}%
which has the property

\begin{equation}
F_{ADB}=-F_{DAB}.  \tag{16}
\end{equation}%
Thus, by using the expressions (15) we find that the field equations (14)
are simplified in the form

\begin{equation}
\partial ^{A}F_{ABD}=\frac{8\pi G_{1+d}}{c^{2}}T_{BD}.  \tag{17}
\end{equation}%
Since%
\begin{equation}
\partial ^{A}F_{ADB}=\partial ^{A}F_{ABD},  \tag{18}
\end{equation}%
the field equations (17) can also be written as

\begin{equation}
\partial ^{A}F_{A(BD)}=\frac{16\pi G_{1+d}}{c^{2}}T_{BD}.  \tag{19}
\end{equation}%
where the bracket $(BD)$ means symmetrization of the indices $B$ and $D.$ It
is worth mentioning that in a $1+3$-dimensional spacetime the field
equations (19) are reduced to the ones proposed by the Novello and Neves [4].

\bigskip\ 

\noindent \textbf{3.- Linearized gravity in a five-dimensional Kaluza-Klein
theory}

\smallskip\ 

In a 5-dimensional spacetime the weak field metric tensor $\gamma _{AB}=\eta
_{AB}+h_{AB},$where $\eta _{AB}=diag(-1,1,1,1,1)$, can be written in the
block-matrix form

\begin{equation}
\gamma _{AB}=\left( 
\begin{array}{cc}
\eta _{\mu \nu }+h_{\mu \nu } & h_{4\mu } \\ 
h_{4\nu } & 1+h_{44}%
\end{array}%
\right) ,  \tag{20}
\end{equation}%
with $\mu ,\nu =0,1,2,3$. If one adopts the Kaluza-Klein ansatz, with

\begin{equation}
h_{\mu \nu }=h_{\mu \nu }(x^{\alpha }),  \tag{21}
\end{equation}%
\begin{equation}
h_{4\mu }=A_{\mu }(x^{\alpha })  \tag{22}
\end{equation}%
and%
\begin{equation}
h_{44}=0,  \tag{23}
\end{equation}%
where $A_{\mu }(x^{\alpha })$\ is identified with the electromagnetic
potential, we discover that the only nonvanishing terms of $\mathcal{F}%
_{DAB} $ are

\begin{equation}
\mathcal{F}_{\mu \nu \alpha }=\frac{1}{2}(h_{\alpha \mu ,\nu }-h_{\alpha \nu
,\mu }),  \tag{24}
\end{equation}%
\begin{equation}
\mathcal{F}_{4\nu \alpha }=\frac{1}{2}\partial _{\nu }A_{\alpha },  \tag{25}
\end{equation}%
and%
\begin{equation}
\mathcal{F}_{\mu \nu 4}=-\frac{1}{2}F_{\mu \nu },  \tag{26}
\end{equation}%
where $F_{\mu \nu }=A_{\nu ,\mu }-A_{\mu ,\nu }$ is the electromagnetic
field strength. Thus, from (15) we find that the nonvanishing components of $%
F_{DAB}$ are

\begin{equation}
F_{\mu \nu \alpha }=\mathcal{F}_{\mu \nu \alpha }+\eta _{\alpha \mu }%
\mathcal{F}_{\nu }-\eta _{\alpha \nu }\mathcal{F}_{\mu },  \tag{27}
\end{equation}%
\begin{equation}
F_{4\nu \alpha }=\frac{1}{2}(\partial _{\nu }A_{\alpha }-\eta _{\alpha \nu
}\partial ^{\beta }A_{\beta }),  \tag{28}
\end{equation}%
\begin{equation}
F_{\mu \nu 4}=-\frac{1}{2}F_{\mu \nu },  \tag{29}
\end{equation}%
and

\begin{equation}
F_{\mu 44}=-\mathcal{F}_{\mu }.  \tag{30}
\end{equation}%
Now, since all fields are independent of the coordinate $x^{4}$ we see that
the field equations (17) can be written as

\begin{equation}
\partial ^{\mu }F_{\mu BD}=\frac{8\pi G_{1+4}}{c^{2}}T_{BD}.  \tag{31}
\end{equation}%
These field equations can be separated as follow

\begin{equation}
\partial ^{\mu }F_{\mu \nu \alpha }=\frac{8\pi G_{1+4}}{c^{2}}T_{\nu \alpha
},  \tag{32}
\end{equation}%
\begin{equation}
\partial ^{\mu }F_{\mu 4\nu }=\frac{8\pi G_{1+4}}{c^{2}}T_{4\nu },  \tag{33}
\end{equation}%
\begin{equation}
\partial ^{\mu }F_{\mu \nu 4}=\frac{8\pi G_{1+4}}{c^{2}}T_{\nu 4},  \tag{34}
\end{equation}%
and%
\begin{equation}
\partial ^{\mu }F_{\mu 44}=\frac{8\pi G_{1+4}}{c^{2}}T_{44}.  \tag{35}
\end{equation}%
Using (28) and (29), we find that (33) and (34) lead exactly to the same
field equations, namely

\begin{equation}
\partial ^{\mu }F_{\nu \mu }=4\pi J_{\nu },  \tag{36}
\end{equation}%
where $J_{\nu }=\frac{4G_{1+4}}{c^{2}}T_{\nu 4}$ is the electromagnetic
current. Of course, the field equations (32) and (36) correspond to
linearized gravity and Maxwell field equations, respectively. If we set $%
T_{44}=0$ then one can see that the field equation (35) is pure gauge
expression. In fact if one assumes the transverse traceless gauge in five
dimensions

\begin{equation}
\begin{array}{c}
h^{AB},_{B}=0, \\ 
\\ 
h=\eta ^{AB}h_{AB}=0,%
\end{array}
\tag{37}
\end{equation}%
one discover that $F_{\mu 44}$ is identically equal to zero.

By completeness let us observe that (37) leads to

\begin{equation}
\begin{array}{c}
h^{\mu \nu },_{\nu }=0, \\ 
\\ 
h=\eta ^{\mu \nu }h_{\mu \nu }=0,%
\end{array}
\tag{38}
\end{equation}%
and the Lorentz gauge for $A_{\mu }$

\begin{equation}
A^{\mu },_{\mu }=0.  \tag{39}
\end{equation}%
Consequently one finds that, in the gauge (38) and (39), the field equations
(32) and (36) reduces to

\begin{equation}
\square ^{2}h_{\mu \nu }=-\frac{16\pi G}{c^{2}}T_{\mu \nu }  \tag{40}
\end{equation}%
and

\begin{equation}
\square ^{2}A^{\mu }=-4\pi J^{\mu },  \tag{41}
\end{equation}%
respectively, where $\square ^{2}=\eta ^{\mu \nu }\partial _{\mu }\partial
_{\nu }$ is the d'Alembertian operator. Thus, we have found a framework in
which the gravitational and electromagnetic waves can be treated in the same
footing.

\bigskip\ 

\noindent \textbf{4.- The De Sitter generalization}

\smallskip\ 

In order to generalize the formalism described in the previous section to a
De Sitter scenario we shall replace the flat metric $\eta _{\mu \nu }$ by
the de Sitter metric $f_{\mu \nu }(x^{\alpha })$. In this case the perturbed
Kaluza-Klein metric $\gamma _{AB}$ takes the form

\begin{equation}
\gamma _{AB}=\left( 
\begin{array}{cc}
f_{\mu \nu }+h_{\mu \nu } & A_{\mu } \\ 
A_{\nu } & 1%
\end{array}%
\right) ,  \tag{42}
\end{equation}%
whose inverse is given by

\begin{equation}
\gamma ^{AB}=%
\begin{pmatrix}
f^{\mu \nu }-h^{\mu \nu } & -A^{\mu } \\ 
-A^{\nu } & 1%
\end{pmatrix}%
.  \tag{43}
\end{equation}

By combining the results from section 2 and appendix B it is not difficult
to obtain the generalized field equations%
\begin{equation}
D^{A}F_{ABD}-\frac{3}{2l^{2}}(h_{BD}-hf_{BD})=\frac{8\pi G_{1+d}}{c^{2}}%
T_{BD}.  \tag{44}
\end{equation}%
Thus, using (42) and considering that in five dimensions $d=4$ and $\Lambda =%
\frac{6}{l^{2}}$ we find

\begin{equation}
D^{\alpha }F_{\alpha \mu \nu }-\frac{\Lambda }{4}(h_{\mu \nu }-hf_{\mu \nu
})=\frac{8\pi G_{1+4}}{c^{2}}T_{\mu \nu },  \tag{45}
\end{equation}%
and%
\begin{equation}
D^{\nu }F_{\mu \nu }-\frac{\Lambda }{4}A_{\mu }=4\pi J_{\mu }.  \tag{46}
\end{equation}%
Here, we have used the fact that $D^{4}F_{4AB}=0$ and $f_{4D}=0.$ Observe
that in this case the electromagnetic field strength $F_{\mu \nu }$ is given
by

\begin{equation*}
F_{\mu \nu }=D_{\mu }A_{\nu }-D_{\nu }A_{\mu }.
\end{equation*}%
The field equations (45) and (46) are remarkable because if we set

\begin{equation}
m^{2}=\frac{\Lambda }{4}  \tag{47}
\end{equation}%
one discovers that, up to factors, both the graviton and the photon have the
same mass $m$ and even more intriguing is that such a mass is proportional
to the square root of the cosmological constant.

\bigskip\ 

\noindent \textbf{5.- Final remarks}

\smallskip\ 

The present work may have a number of interesting developments. In
particular, as Novello and Neves [4] have shown, the equation (15) can be
derived from the formula

\begin{equation}
\partial ^{A\ast }F_{A(BD)}=0,  \tag{48}
\end{equation}%
where

\begin{equation}
^{\ast }F^{A(BD)}=\varepsilon ^{ABEF}F_{EF}^{D}+\varepsilon
^{ADEF}F_{EF}^{B},  \tag{49}
\end{equation}%
with $\varepsilon ^{ABEF}$ the completely antisymmetric symbol. Thus, one
may consider an alternative approach [5] for duality aspects of linearized
gravity [6] (see Refs. [7]--[11]) as in the case of Maxwell theory (see
Refs. [12]-[13]).

Another source of physical interest of the present formalism is a possible
connection with the Randall-Sundrum brane world scenario [14]-[15], with
gravitational waves formalism (see [16] and references there in) and with
quantum linearized gravity [17]-[18]. Moreover, our work may be also useful
to clarify some aspects about the relation between the mass of the graviton
and the cosmological constant, which has been subject of some controversy
[19]-[20].

Aside from theoretical developments, the present work opens also the
possibility to make a number of applications for linearized gravity arising
from the Maxwell theory itself. Let just outline one possibility. In Maxwell
theory the concept of polarization scattering is of considerable interest in
optical physics (see Ref. [21] and references therein). The subject of
interest in this arena is to describe in a complex setting the interaction
of polarized waves with a target. It turns out that the useful mathematical
tool in the scattering radiation process is the so-called Mueller matrix
[22] (see also Ref. [23] and references therein). What seems to be
interesting about such a matrix is that its elements refer to intensity
measurements only. Let us recall the main ideas for the construction of the
Mueller matrix.

A polarized radiation field may be represented by a 2-dimensional complex
field

\begin{equation}
\left( 
\begin{array}{c}
E_{x} \\ 
E_{y}%
\end{array}%
\right) .  \tag{50}
\end{equation}%
From the components of this vector one may define the Hermitian coherency
matrix

\begin{equation}
\mathbf{C}=\left( 
\begin{array}{cc}
E_{x}E_{x}^{\ast } & E_{x}E_{y}^{\ast } \\ 
E_{x}^{\ast }E_{y} & E_{y}E_{y}^{\ast }%
\end{array}%
\right)  \tag{51}
\end{equation}%
and the vector

\begin{equation}
\mathbf{g}=\left( 
\begin{array}{c}
g_{0} \\ 
g_{1} \\ 
g_{2} \\ 
g_{3}%
\end{array}%
\right) ,  \tag{52}
\end{equation}%
where

\begin{equation}
g_{0}=E_{x}E_{x}^{\ast }+E_{y}E_{y}^{\ast },  \tag{53}
\end{equation}%
\begin{equation}
g_{1}=E_{x}E_{x}^{\ast }-E_{y}E_{y}^{\ast },  \tag{54}
\end{equation}

\begin{equation}
g_{2}=E_{x}^{\ast }E_{y}+E_{x}E_{y}^{\ast },  \tag{55}
\end{equation}%
and%
\begin{equation}
g_{3}=i(E_{x}^{\ast }E_{y}-E_{x}E_{y}^{\ast }).  \tag{56}
\end{equation}

The Jones matrix $J$ and the Mueller matrix $M$ apply to $\mathbf{c}$ and
the vector $\mathbf{g,}$ respectively as follows

\begin{equation}
\mathbf{C}^{\prime }=\mathbf{JC}  \tag{57}
\end{equation}%
and%
\begin{equation}
\mathbf{g}^{\prime }\mathbf{=Mg.}  \tag{58}
\end{equation}%
It turns out that $\mathbf{J}$ can be identified with a SU(2) matrix, while $%
\mathbf{M}$ is a 4$\times $4 matrix augmented form of $O_{3}$. Of course the
matrices $\mathbf{J}$ and $\mathbf{M}$ must be related,

\begin{equation}
M=\frac{1}{2}TrJ^{\ast T}\sigma J\sigma ,  \tag{59}
\end{equation}%
where $\sigma $ denotes the four Pauli matrices (see Ref. [22] for details).

Let us make the identification%
\begin{equation}
F_{0i\alpha }=E_{i\alpha }.  \tag{60}
\end{equation}%
Here $F_{0i\alpha }$ denotes some of the components of $F_{ADB}$ according
to the expression (15). The idea is now to consider the generalization of
(50)

\begin{equation}
\left( 
\begin{array}{c}
E_{x\alpha } \\ 
E_{y\alpha }%
\end{array}%
\right)  \tag{61}
\end{equation}%
and to consider the analogue of (53)-(56), namely

\begin{equation}
G_{0}=E_{x}^{\alpha }E_{x\alpha }^{\ast }+E_{y}^{\alpha }E_{y\alpha }^{\ast
},  \tag{62}
\end{equation}%
\begin{equation}
G_{1}=E_{x}^{\alpha }E_{x\alpha }^{\ast }-E_{y}^{\alpha }E_{y\alpha }^{\ast
},  \tag{63}
\end{equation}

\begin{equation}
G_{2}=E_{x}^{\ast \alpha }E_{y\alpha }+E_{x}^{\alpha }E_{y\alpha }^{\ast } 
\tag{64}
\end{equation}%
and%
\begin{equation}
G_{3}=i(E_{x}^{\ast \alpha }E_{y\alpha }-E_{x}^{\alpha }E_{y\alpha }^{\ast
}).  \tag{65}
\end{equation}%
In this way, we can apply the Mueller matrix $M$ as in (58); $\mathbf{G}%
^{\prime }\mathbf{=MG}$. In fact, since $\mathbf{M}$ contains the
information of the intensities of both gravitational and electromagnetic
radiation via the quantity $F_{0i\alpha }$ one should expect a broad kind of
applications of $\mathbf{M}$ in a gravitational wave context.

\bigskip\ 

\noindent \textbf{6.- Appendix A}

\smallskip\ 

In this appendix, we shall briefly review the De Sitter (anti-De Sitter)
space, which provides us with one of the simplest solutions of Einstein
field equations with cosmological constant. For this purpose let us consider
the line element%
\begin{equation}
ds^{2}=\eta _{AB}dx^{A}dx^{B}+(dx^{d+1})^{2},  \tag{66}
\end{equation}%
along with the constraint

\begin{equation}
\eta _{AB}x^{A}x^{B}+(x^{d+1})^{2}=l^{2},  \tag{67}
\end{equation}%
with $l$\ being a constant, the indices $A,B$ running from zero to $(1+d)-1$%
, and $\eta _{AB}$ being a flat metric.

From Eq. (67) we get

\begin{equation}
(dx^{d+1})^{2}=\frac{x_{A}x_{B}dx^{A}dx^{B}}{(l^{2}-x_{C}x^{C})}.  \tag{68}
\end{equation}%
Thus, Eq. (66) becomes

\begin{equation}
ds^{2}=f_{AB}dx^{A}dx^{B},  \tag{69}
\end{equation}%
where the metric $f_{AB}$\ is given by

\begin{equation}
f_{AB}=\eta _{AB}+\frac{x_{A}x_{B}}{l^{2}-x_{C}x^{C}},  \tag{70}
\end{equation}%
The inverse of $f_{AB}$ is found to be

\begin{equation}
f^{AB}=\eta ^{AB}-\frac{x^{A}x^{B}}{l^{2}-x_{C}x^{C}},  \tag{71}
\end{equation}

Using (70) and (71) one discovers that the Christoffel symbols and the
Riemann tensor are

\begin{equation}
\Gamma _{CD}^{A}=\frac{x^{A}}{l^{2}}f_{AB}  \tag{72}
\end{equation}%
and

\begin{equation}
R_{ABCD}=\frac{1}{l^{2}}(f_{AC}f_{BD}-f_{AD}f_{BC}),  \tag{73}
\end{equation}%
respectively. Consequently, we find that $f_{AB}$ is a solution of the
vacuum Einstein field equations with cosmological constant $\Lambda ,$

\begin{equation}
R_{AB}-\frac{1}{2}f_{AB}R+\Lambda f_{AB}=0,  \tag{74}
\end{equation}%
provided the constants $\Lambda $\ and $l^{2}$ are related by

\begin{equation}
\Lambda =\frac{(D-2)(D-1)}{2l^{2}},  \tag{75}
\end{equation}%
where $D=1+d$. Observe that in $D=4$ (75) implies that $\Lambda =\frac{3}{%
l^{2}}.$ It is interesting to note that these results are independent of the
signature of $\eta _{AC}.$

\bigskip\ 

\noindent \textbf{7.- Appendix B}

\smallskip\ 

In this appendix we shall study the linearized gravity with $f_{AB}$\ as a
metric background. Consider the perturbed metric

\begin{equation}
\gamma _{AB}=f_{AB}+h_{AB},  \tag{76}
\end{equation}%
where $h_{AB}$ is a small perturbation, that is $\left\vert
h_{AB}\right\vert \ll 1.$ To first order in $h_{AB}$ one finds

\begin{equation}
\gamma ^{AB}=f^{AB}-h^{AB}.  \tag{77}
\end{equation}

The Christoffel symbols can be written in the interesting form

\begin{equation}
\Gamma _{CD}^{A}=\Omega _{CD}^{A}+H_{CD}^{A},  \tag{78}
\end{equation}%
where $\Omega _{CD}^{A}$\ are the Christoffel symbols associated with $%
f_{AB} $ and $H_{CD}^{A}$ is defined by%
\begin{equation}
H_{CD}^{A}=\frac{1}{2}f^{AE}(D_{C}h_{DE}+D_{D}h_{CE}-D_{E}h_{CD}).  \tag{79}
\end{equation}%
Here, the symbol $D_{A}$ denotes covariant derivatives with $\Omega
_{CD}^{A} $ as a connection.

By using (78) it is straightforward to check that the Riemann tensor can be
written in the form

\begin{equation}
R_{BCD}^{A}=\mathcal{R}_{BCD}^{A}+D_{C}H_{BD}^{A}-D_{D}H_{BC}^{A},  \tag{80}
\end{equation}%
where $\mathcal{R}_{BCD}^{A}$\ is the Riemann tensor associated with $\Omega
_{CD}^{A}$.

With the help of the commutation relation

\begin{equation}
D_{C}D_{D}h_{AB}-D_{D}D_{C}h_{AB}=-\mathcal{R}_{ACD}^{E}h_{EB}-\mathcal{R}%
_{BCD}^{E}h_{AE}  \tag{81}
\end{equation}%
we find that the Riemann curvature tensor (80) can also be written as%
\begin{equation}
R_{ABCD}=D_{A}\mathcal{F}_{CDB}-D_{B}\mathcal{F}_{CDA}+\mathcal{R}_{ABCD}+%
\frac{1}{2}\mathcal{R}_{DAB}^{E}h_{EC}-\frac{1}{2}\mathcal{R}%
_{CAB}^{E}h_{ED}.  \tag{82}
\end{equation}%
Here, the symbol $\mathcal{F}_{CDB}$ takes the form

\begin{equation}
\mathcal{F}_{CDB}=\frac{1}{2}(D_{D}h_{BC}-D_{C}h_{BD}).  \tag{83}
\end{equation}%
Observe that $\mathcal{F}_{CDB}$ can be obtained from (9) by replacing
ordinary partial derivatives by covariant derivatives$.$ From (82) we get
the Ricci tensor

\begin{equation}
R_{BD}=D^{A}\mathcal{F}_{ADB}+D_{B}\mathcal{F}_{D}+\mathcal{R}_{BD}-\frac{1}{%
2}\mathcal{R}_{ABCD}h^{AC}+\frac{1}{2}\mathcal{R}_{EB}h_{D}^{E}  \tag{84}
\end{equation}%
where

\begin{equation}
\mathcal{F}_{A}=f^{CB}\mathcal{F}_{ACB}.  \tag{85}
\end{equation}%
Thus, we find that the Ricci scalar $R$ is given by

\begin{equation}
R=2D^{A}\mathcal{F}_{A}+\mathcal{R-R}_{BD}h^{BD}.  \tag{86}
\end{equation}

Substituting (84) and (86)\ into the Einstein weak field equations in $1+d$
dimensions with cosmological constant

\begin{equation}
R_{BD}-\frac{1}{2}(f_{BD}+h_{BD})R+(f_{BD}+h_{BD})\Lambda =\frac{8\pi G_{1+d}%
}{c^{2}}T_{BD,}  \tag{87}
\end{equation}%
we obtain

\begin{equation}
\begin{array}{c}
D^{A}\mathcal{F}_{ADB}+D_{B}\mathcal{F}_{D}-f_{BD}D^{A}\mathcal{F}_{A}-\frac{%
1}{2}\mathcal{R}_{ABCD}h^{AC} \\ 
\\ 
+\frac{1}{2}\mathcal{R}_{EB}h_{D}^{E}+\frac{1}{2}f_{BD}\mathcal{R}%
_{EF}h^{EF}-\frac{1}{2}h_{BD}\mathcal{R+}h_{BD}\Lambda =\frac{8\pi G_{1+d}}{%
c^{2}}T_{BD}.%
\end{array}
\tag{88}
\end{equation}%
Here, we used the fact that

\begin{equation}
\mathcal{R}_{BD}-\frac{1}{2}f_{BD}\mathcal{R+}f_{BD}\Lambda =0.  \tag{89}
\end{equation}

Since we have

\begin{equation}
\mathcal{R}_{ABCD}=\frac{1}{l^{2}}(f_{AC}f_{BD}-f_{AD}f_{BC})  \tag{90}
\end{equation}%
and

\begin{equation}
\Lambda =\frac{d(d-1)}{2l^{2}},  \tag{91}
\end{equation}%
we discover that (88) is reduced to

\begin{equation}
\begin{array}{c}
D^{A}\mathcal{F}_{ADB}+D_{B}\mathcal{F}_{D}-f_{BD}D^{A}\mathcal{F}_{A} \\ 
\\ 
-\frac{1}{2l^{2}}(d-1)(h_{BD}-hf_{BD})=\frac{8\pi G_{1+d}}{c^{2}}T_{BD}.%
\end{array}
\tag{92}
\end{equation}%
Thus, by defining the symbol%
\begin{equation}
F_{ADB}\equiv \mathcal{F}_{ADB}+f_{BA}\mathcal{F}_{D}-f_{BD}\mathcal{F}_{A},
\tag{93}
\end{equation}%
we find that the field equations (92) are simplified in the form

\begin{equation}
D^{A}F_{ABD}-\frac{1}{2l^{2}}(d-1)(h_{BD}-hf_{BD})=\frac{8\pi G_{1+d}}{c^{2}}%
T_{BD}.  \tag{94}
\end{equation}

In four dimensions $d=3$ and $\Lambda =\frac{3}{l^{2}}$. Therefore (94)
becomes

\begin{equation}
D^{A}F_{ABD}-\frac{\Lambda }{3}(h_{BD}-hf_{BD})=\frac{8\pi G_{1+d}}{c^{2}}%
T_{BD},  \tag{95}
\end{equation}%
which are the field equations obtained by Novello and Neves [4].

\bigskip\ 

\noindent \textbf{Acknowledgments: }We would like to thank C. M. Yee and E.
A. Le\'{o}n for helpful comments. This work was supported in part by the UAS
under the program PROFAPI-2006.

\smallskip\


\begin{thebibliography}{99}
\bibitem{1} J. A. Nieto, Mod. Phys. Lett. A \textbf{20}, 135 (2005);
hep-th/0311083.

\bibitem{2} G. 't Hooft, L. Susskind, E. Witten, M. Fukugita, L. Randall, L.
Smolin, J. Stachel, C. Rovelli, G. Ellis, S. Weinberg, R. Penrose, Nature 
\textbf{433}, 257 (2005).

\bibitem{3} M.J. Duff, "Kaluza-Klein theory in perspective", talk given at
The Oskar Klein Centenary Symposium, Stockholm, Sweden, 19-21 Sep 1994. In
*Stockholm 1994, The Oskar Klein centenary* 22-35, and Cambridge U. -
NI-94-015 (94/10,rec.Oct.) 38 p. Texas A\&M U. College Station -
CTP-TAMU-94-022 (94/10) 38 p; hep-th/9410046.

\bibitem{4} M. Novello and R. P. Neves, Class. Quant. Grav. \textbf{20}, L67
(2003); "Apparent mass of the graviton in a de Sitter background",
gr-qc/0210058.

\bibitem{5} E. A. Le\'{o}n and J. A. Nieto, work in progress (2006).

\bibitem{6} J. A. Nieto, Phys. Lett. A \textbf{262}, 274 (1999);\
hep-th/9910049.

\bibitem{7} M. Henneaux and C. Teitelboim, Phys. Rev. D \textbf{71}, 024018
(2005); gr-qc/0408101.

\bibitem{8} C. M. Hull, JHEP \textbf{0109}, 027 (2001); hep-th/0107149.

\bibitem{9} S. Cnockaert, "Higher spin gauge field theories: Aspects of
dualities and interactions", (Brussels U., PTM), Jun 2006. 193pp., Ph.D.
Thesis, e-Print Archive: hep-th/0606121.

\bibitem{10} A. J. Nurmagambetov, "Duality-symmetric approach to general
relativity and supergravity", To memory Vladimir G. Zima and Anatoly I.
Published in SIGMA \textbf{2}, 020,2006;\ hep-th/0602145.

\bibitem{11} B. Julia, J. Levie and S. S. Ray, JHEP \textbf{0511}, 025
(2005);\ hep-th/0507262.

\bibitem{12} E. Witten, Selecta Math.\textbf{1}, 383 (1995); hep-th/9505186.

\bibitem{13} Y. Lozano, Phys. Lett. B 364, 19 (1995); hep-th/9508021.

\bibitem{14} L. Randall and R. Sundrum, Phys. Rev. Lett. \textbf{83}, 4690
(1999).

\bibitem{15} S. B. Giddings, E. Katz and L. Randall, JHEP \textbf{0003}, 023
(2000); hep-th/0002091; H. Collins and B. Holdom, Phys. Rev. \textbf{D62},
124008 (2000); hep-th/0006158;\ N. Deruelle and T. Dolezel, Phys. Rev. 
\textbf{D 64}, 103506 (2001): gr-qc/0105118

\bibitem{16} J. A. Nieto, J. Saucedo and V.M. Villanueva, Phys. Lett. 
\textbf{A 312}, 175 (2003): hep-th/0303123.

\bibitem{17} T. Fukai and K. Okano, Prog. Theor. Phys.\textbf{73}, 790 (1985)

\bibitem{18} J.B. Hartle, Phys. Rev. \textbf{D 29}, 2730 (1984).

\bibitem{19} M. Novello and R. P. Neves, Class. Quant. Grav. \textbf{20},
L67 (2003); M. Novello, S. E. Perez Bergliaffa, R. P. Neves, "Replay
acausality of massive charged spin-2 field";\ gr-qc/0304041.

\bibitem{20} S. Deser, A. Waldron, "Acausality of massive charged spin-2
field", hep-th/0304050.

\bibitem{21} S: R. Cloude, Optik \textbf{75}, 26 (1986).

\bibitem{22} S. Huard, \textit{Polarization of Light}, (John Wiley \& Sons
1997).

\bibitem{23} G. Atondo-Rubio, R. Espinosa-Luna and A. Mendoza-Su\'{a}rez,
Opt. Commun. \textbf{244}, 7 (2005).
\end{thebibliography}
\end{document}